\documentclass[prl,twocolumn,superscriptaddress,showpacs,floats,floatfix]{revtex4}

\usepackage{epsfig}
\begin{document}

\draft

\title{Atomic Motion in Single H$_{2}$ and D$_{2}$ Molecule Junction Induced by Phonon Excitation}

\author{Manabu Kiguchi}
\thanks{Present address: Department of Chemistry, Graduate School of Science and Engineering, 
Tokyo Institute of Technology 2-12-1 W4-10 Ookayama, Meguro-ku, Tokyo 152-8551, Japan}
 \affiliation{Division of Chemistry, Graduate School of Science, Hokkaido University, N10W8, Kita, Sapporo, 060-0810, Japan}
 \affiliation{PRESTO, Japan Science and Technology Agency, N10W8, Kita, Sapporo, 060-0810, Japan}

\author{Tomoka Nakazumi}
\thanks{Present address: Department of Chemistry, Graduate School of Science and Engineering, 
Tokyo Institute of Technology 2-12-1 W4-10 Ookayama, Meguro-ku, Tokyo 152-8551, Japan}
\affiliation{Division of Chemistry, Graduate School of Science, Hokkaido University, N10W8, Kita, Sapporo, 060-0810, Japan}

\author{Kunio Hashimoto}
\affiliation{Division of Chemistry, Graduate School of Science, Hokkaido University, N10W8, Kita, Sapporo, 060-0810, Japan}

\author{Kei Murakoshi}
\affiliation{Division of Chemistry, Graduate School of Science, Hokkaido University, N10W8, Kita, Sapporo, 060-0810, Japan}

\date{\today}

\begin{abstract}
We have investigated Au atomic contacts in H$_{2}$ and D$_{2}$ environment by conductance measurement 
and $dI/dV$ spectroscopy. A single H$_{2}$ or D$_{2}$ molecule was found to bridge Au electrodes. In the case 
of the Au/H$_{2}$/Au junction, symmetric peaks were observed in $dI/dV$ spectra, while they were not 
observed for the Au/D$_{2}$/Au junction. The shape of the peaks in $dI/dV$ spectra originated from the 
structural change of the single molecule junction induced by the phonon excitation. The structural 
change could occur only for the Au/H$_{2}$/Au junction. The difference in the two single molecule 
junctions could be explained by larger zero point energy of Au-H$_{2}$ vibration mode than that in the 
Au/H$_{2}$/Au junction. 
\end{abstract}

\maketitle

\section{INTRODUCTION}
\label{sec1}

Nano junctions using single molecules have attracted wide attention in recent year \cite{1,2}. In 
order to investigate the atomic scale single molecular junctions, it is important to characterize a 
single molecule bridging between metal electrodes using some spectroscopic measurements. As for 
the spectroscopy of the single molecule junctions, point contact spectroscopy (PCS) \cite{3,4,5}, inelastic 
tunneling spectroscopy (IETS) \cite{6,7,8,9,10}, and action spectroscopy have been investigated for the single 
molecule junctions \cite{11}. In these spectroscopies, electron current through the single molecule 
junction is measured as a function of bias voltage. In the IETS and PCS, spectrum is obtained as an 
inelastic scattering event of electrons with sufficient energy to excite certain vibration mode. The 
vibration energy is determined from the bias voltage, in which conductance of the single molecule 
junction abruptly changes. The difference between PCS and IETS is operating conductance regime. 
IETS operates in low conductance regime ($G$ $\ll$1 $G_{0}$, where $G_{0}$ =2$e^{2}/h$), while PCS operates in 
conductance regime close to 1 $G_{0}$ \cite{5,12,13}. The PCS and IETS have been applied for various 
single molecule junctions with H$_{2}$, C$_{60}$, benzene, alkane dithiol \cite{3,4,5,6,7,8,9}.
 
Action spectroscopy of the single molecule junction was a recently developed technique to 
detect the vibration modes of the single molecule junctions \cite{11}. The action spectroscopy observes 
the abrupt conductance change (peak in $dI/dV$ spectra) originated from the structural change of the 
single molecule junction induced by the phonon excitation. The energy of the peak provides the 
vibration energy of the single molecular junction. The action spectroscopy of the single molecular 
junction was applied for the single molecular junctions with H$_{2}$, O$_2$, and CO \cite{11}. In the case of 
single molecules adsorbed on metal surfaces, the action spectroscopy has some advantage compared 
to IETS \cite{14,15,16}. It can detect the vibration modes which could not be detected by IETS. In addition, 
action spectroscopy can be applied for the system, in which IETS signal is too weak to detect, since 
the signal of the action spectra is much larger than that of IETS. The action spectroscopy has been 
applied for various molecules on metal substrates. On the other hand, there is little study of the 
action spectroscopy of the single molecular junction, and thus, the character of this spectroscopy is 
not clear up to now. It is important to investigate the characters of the action spectroscopy of the 
single molecular junction. In the present study, the action spectroscopy has been performed for the 
Au atomic contact in H$_{2}$ and D$_{2}$ environment. This system has been already investigated by the 
conductance measurement, $dI/dV$ spectra, and theoretical calculation by other groups \cite{17,18,19,20}. 
However, IETS or PCS of this system has not been obtained up to now, and the energy of the 
vibration mode was not fixed yet. The structure of the Au atomic contact in H$_{2}$ environment was not 
defined. In the present study, the atomic configuration of the Au atomic contact in H$_{2}$ and D$_{2}$ 
environment was investigated by the conductance measurement, action spectroscopy. Here it should 
be noticed that hydrogen has large zero point energy due to its small mass. Since the action 
spectroscopy detects the structural change of the single molecular junctions induced by the phonon 
excitation, it provides the information of the atomic motion of the single molecular junction. The 
effect of the zero point motion on the atomic scale structural change of the single molecule junction 
could be revealed by the action spectroscopy. In the present study, we have investigated the effect of 
the zero point motion on the atomic motion of the single molecular junction by investigation of the 
action spectroscopy of the Au atomic contacts in H$_{2}$ and D$_{2}$ environment.

\section{EXPERIMENTAL}
\label{sec2}

The measurements have been performed using the mechanically controllable break 
junction (MCBJ) technique (see Ref. \cite{2} for a detailed description). A notched Au wire (0.1 mm in 
diameter, 10 mm in length) was fixed with epoxy adhesive (Stycast 2850FT) on top of a bending 
beam and mounted in a three-point bending configuration inside a vacuum chamber. In ultra high 
vacuum at 4 K, the Au wire was broken by mechanical bending of the substrate, and clean fracture 
surfaces were exposed. The bending could be relaxed to form atomic-sized contacts between the 
wire ends using a piezo element for fine adjustment. H$_{2}$ or D$_{2}$ gas was admitted to the contacts via a 
capillary. DC two-point voltage-biased conductance measurements were performed by applying a 
bias voltage in the range from 10 to 300 mV. Every statistical data set was built from a large 
number (over 3000) of individual digitized conductance traces. AC voltage bias conductance 
measurements were performed using a standard lock-in technique. The conductance was recorded 
for fixed contact configuration using an AC modulation of 1 mV amplitude and a frequency of 
7.777 kHz, while slowly ramping the DC bias between -100 and +100 mV.

\section{RESULTS AND DISCUSSION}
\label{sec3}

\begin{figure}
\begin{center}
\leavevmode\epsfysize=50mm \epsfbox{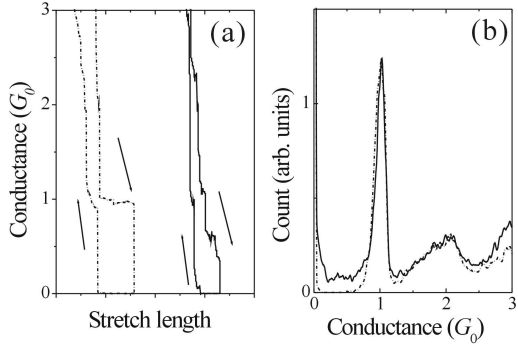}
\caption{(a) Conductance trace and (b) conductance histogram of Au contacts before (dotted line) 
and after (line) introduction of H$_{2}$ at the bias voltage of 100 mV.} \label{fig1}
\end{center}
\end{figure}

Figure~\ref{fig1} shows typical conductance traces and histograms for Au contacts before and after 
introduction of H$_{2}$. The stretch length was the displacement of the distance between the stem parts 
of the Au electrodes which were fixed on the substrate. Before introduction of H$_{2}$, 
conductance decreased in a stepwise fashion, with each step occurring at integer multiples of $G_{0}$. 
The corresponding conductance histograms showed a peak near 1 $G_{0}$, which corresponded to a 
clean Au atomic contact. After introduction of H$_{2}$, the conductance decreased multi-stepwise 
fashion below 1 $G_{0}$. The widths of the multiple steps were not well-defined, and showed various 
value less than 1 $G_{0}$. The corresponding conductance histogram showed a broad feature below 1 $G_{0}$. 
The appearance of steps and features below 1 $G_{0}$ agreed with the previously reported results \cite{17}. 
Here, it should be noticed that the return conductance traces showed the steps around 0.1-0.3 $G_{0}$, 
before making large contacts. The certain atomic configurations showing conductance values below 
1 $G_{0}$ would be formed in both breaking and making contacts.

\begin{figure}
\begin{center}
\leavevmode\epsfysize=50mm \epsfbox{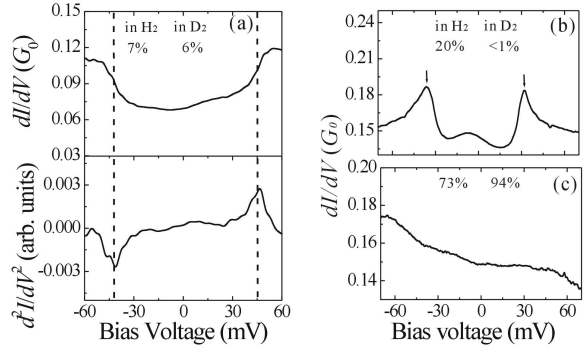}
\caption{
Schematic $dI/dV$ spectra of Au contacts in H$_{2}$ environment: (a) normal, (b) peak, (c) no 
symmetric feature type spectra. The percentage of the normal, peak, non symmetric type $dI/dV$ 
spectra for single Au/H$_{2}$/Au and Au/D$_{2}$/Au junctions is shown in the figure. The number of the total 
$dI/dV$ spectra was 200.
} \label{fig2}
\end{center}
\end{figure}

The differential conductance ($dI/dV$) spectra were measured for the Au atomic contact in 
H$_{2}$ environment at a conductance value of 0.1-0.5 $G_{0}$. Three types of $dI/dV$ spectra were observed, 
as shown in Fig.~\ref{fig2}. The first "normal" spectrum (Fig.~\ref{fig2}(a)) showed an increase in the differential 
conductance symmetrically around 45 meV, and clear symmetric peaks were observed in the second 
derivative ($dI^{2}/dV^{2}$). The second "peak" spectrum (Fig.~\ref{fig2}(b)) showed symmetrical peaks, and the 
third "non symmetric" spectrum (Fig.~\ref{fig2}(c)) did not show any symmetric features. The third "non 
symmetric" spectra were most frequently observed for the Au contact in H$_{2}$ and D$_{2}$ environment. 
The increase in conductance observed for the "normal" spectra could be explained by 
the phonon excitation at the single molecular junction \cite{3,4,5,6,7,8,9,12,13}. The 
single-channel model predicts conductance enhancement below a transmission 
probability of 0.5 and suppression of conductance above this. In the present study, the 
spectra were measured for the junction having zero bias conductance below 0.5 
$G_{0}$. The increase in the conductance agreed with the theoretical prediction. In the 
single molecular junction having low conductance, an additional tunneling channel for 
electrons was opened, when the bias voltage was increased and crossed the threshold for 
excitation of a vibration mode.
This opening of the new channel resulted in a sudden increase in the 
differential conductance at the threshold voltage. Here, it should be noticed that $dI/dV$ spectra of the 
Au atomic contacts in H$_{2}$ environment has not been obtained before \cite{17,18,19,20}. In the present study, 
we could measure the vibration spectra of this system by measuring the carefully prepared sample. 
The peak observed in the "peak" spectra could be explained by the abrupt switching between two 
slightly different local geometric configurations induced by the phonon excitation \cite{11}. Action 
spectroscopy observes these peaks to determine the energy of the vibration mode. 
The non symmetric feature in the "non symmetric" spectra could be explained by 
interference of electron waves which scatter on defects or impurities close to the contact 
\cite{2,3}. Since the distribution of the defects or impurities and shape of the left and 
right electrodes would be different from each other, that is, non symmetric, the non 
symmetric feature could appear in the spectra. The conductance fluctuation 
is most enhanced for the contact showing a conductance value around 0.5 $G_{0}$, if the conductance 
channel is a single one channel \cite{2}. Therefore, the "non-symmetric" spectra were most frequently 
observed for the Au contact in H$_{2}$ and D$_{2}$ environment showing conductance of 0.1-0.5 $G_{0}$. 

\begin{figure}
\begin{center}
\leavevmode\epsfysize=50mm \epsfbox{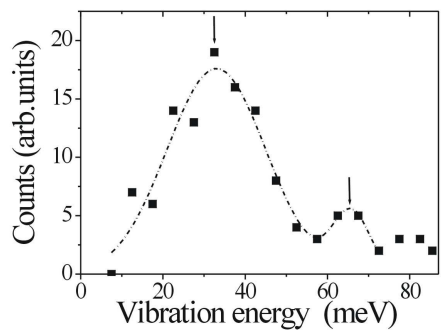}
\caption{
Distribution of phonon energy for the Au/H$_{2}$/Au junctions obtained from the action 
spectroscopy of the single molecule junction.} \label{fig3}
\end{center}
\end{figure}

In order to determine the vibration energy from action spectra, 122 differential 
conductance spectra showing peaks in $dI/dV$ spectra were collected for junctions having 
conductance of 0.05-0.4 $G_{0}$. Figure~\ref{fig3}(b) shows the distribution of vibration energy. The broad 
features were observed around 33 and 66 meV. The two modes were also observed for the "normal" 
$dI/dV$ spectra. The vibration mode of 33 meV was observed in Fig.~\ref{fig2}(a). Since the phonon energy of 
the Au atomic contact is around 10-20 meV \cite{2}, the observed vibration modes would be the 
vibration modes of the bonding between Au and hydrogen. The distribution of the vibration energy 
could not be defined for the Au contact in D$_{2}$ environment due to the difficulty in obtaining the 
"peak" spectra.

The structure of the Au atomic contact in H$_{2}$ environment is discussed based on the present 
experimental results and previously reported theoretical calculation result \cite{19}. After introduction of 
H$_{2}$, the Au contact showed a conductance value below 1 $G_{0}$. The vibration modes between Au and 
hydrogen were observed around 33 and 66 meV in the action spectra. The interaction between Au 
atomic contact with H atom or H$_{2}$ molecule have been investigated by DFT calculation \cite{19}. H 
atom and H$_{2}$ molecule were stably incorporated into the Au atomic contact. Because of the 
incorporation of hydrogen into the contact, the conductance of the Au atomic contact decreased 
from 1 $G_{0}$ to 0.6-0.01 $G_{0}$, depending on the atomic configuration of the contact. The decrease in 
conductance was possibly due to scattering or interference of conducting electrons in the contact. 
No preferential atomic configurations were found for the hydrogen incorporated Au atomic contact. 
Although there was no clear difference in conductance between the Au atomic contact with a single 
H atom and H$_{2}$ molecule, there was clear difference in the vibration energy between two contacts 
\cite{20}. The Au atomic contact with a H atom showed the vibration mode around 150-220meV 
corresponding to the vibration of the H atom along the contact axis. Comparatively, the Au atomic 
contact with a H$_{2}$ molecule showed two transverse modes of the H$_{2}$ molecule around 25-150 meV, 
in addition to the internal H$_{2}$ stretching mode around 180-250meV. These experimental and 
theoretical calculation results indicated that a H$_{2}$ molecule bridged Au electrodes in the present 
experimental condition. 
The formation of the Au/H$_{2}$/Au junction was supported by the conductance trace. 
In the return conductance trace, the conductance jumped to a conductance value of 
0.1-0.4 $G_{0}$. Before making the contact, H$_{2}$ molecules would adsorb on Au 
electrodes without dissociation. In making the contact, a single H$_{2}$ molecule 
would probably bridge Au electrodes in the initial stage. The vibration mode was also 
observed around 30-50meV for the contact having conductance of 0.1-0.4 $G_{0}$, 
which was formed in making the contact. The close agreement in conductance and 
vibration energy supported that the atomic contact (conductance: 0.1-0.4 $G_{0}$, 
vibration energy: 40 meV) formed in breaking the contact would be similar to that in 
making the contact, and that a single H$_{2}$ molecule would bridge Au electrodes. 
Now, the Au atomic contact in H$_{2}$ environment was well characterized by the 
conductance measurement, action spectra of the single molecule junction, and 
theoretical calculation. Briefly, the effect of surrounding hydrogen molecules on the 
Au/H$_{2}$/Au junction is discussed. In the present experimental condition, hydrogen 
molecules would adsorb on the Au/H$_{2}$/Au junction. Since the interaction between 
the hydrogen molecule and the hydrogen molecule in the Au/H$_{2}$/Au junction 
would be small, the conductance and vibration energy of the Au/H$_{2}$/Au junction 
would be insensitive to the presence of the surrounding hydrogen.

Next, $dI/dV$ spectra of the Au/H$_{2}$/Au and Au/D$_{2}$/Au junctions were discussed. Three types 
of spectra ("normal", "peak" and "non symmetric") were observed in $dI/dV$ spectra. There was clear 
difference of the distribution of spectra between the Au/H$_{2}$/Au and Au/D$_{2}$/Au junctions. The 
percentage of the "normal", "peak" and "non symmetric" spectra were 7 $\%$. 20 $\%$, 73 $\%$ for the 
Au/H$_{2}$/Au junction, and 6 $\%$, $<$ 1$\%$, and 94 $\%$ for the Au/D$_{2}$/Au junction, respectively. The "peak" 
spectra (action spectra) were frequently observed for the Au/H$_{2}$/Au junction, while they were not 
observed for the Au/D$_{2}$/Au junction.

\begin{figure}
\begin{center}
\leavevmode\epsfysize=35mm \epsfbox{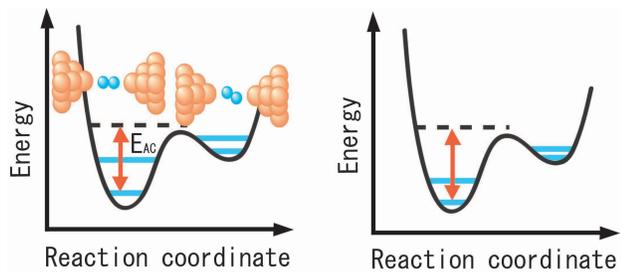}
\caption{
Potential energy surface of Au/H$_{2}$/Au and Au/D$_{2}$/Au junctions.} \label{fig4}
\end{center}
\end{figure}

The peak in the $dI/dV$ spectra is discussed using a model of vibrationlly induced two level 
systems (see Fig.~\ref{fig4}) \cite{11}. In this model, the potential curve of the molecular junction is represented 
as the double well potential with ground states $\Psi_{1}$ and meta stable $\Psi_{2}$ in the two energy minima. 
The two energy minima are separated by the activation barrier ($E_{AC}$). The molecular junction could 
be vibrationally excited by the conduction electron. If the junction is fully exited, the junction with 
ground state $\Psi _{1}$ could overcome the activation barrier, and change into the junction with meta stable 
$\Psi _{2}$, which leads to the abrupt change in conductance (peak in the $dI/dV$ spectra). The $\Psi _{1}$ and $\Psi _{2}$ 
states have slightly different local geometrical configurations, such as adsorption site of a H$_{2}$ 
molecule to Au electrodes, molecule tilt angle, configuration of the Au electrodes. Appearance of 
the "peak" spectra only for the Au/H$_{2}$/Au junction indicated that the structural change could occur 
only for the Au/H$_{2}$/Au junction. 

The difference between the two single molecule junctions is discussed by considering the 
zero point energy of the single molecular junction. Since a H$_{2}$ molecule has half the mass of a D$_{2}$ 
molecule, the energy of Au-H$_{2}$ vibration mode and zero point motion would be larger than those for 
D$_{2}$. On the other hand, the potential curve would be the same for both single molecular junctions 
\cite{15}. Therefore, $E_{AC}$ would be smaller for the Au/H$_{2}$/Au junction than that for the Au/D$_{2}$/Au 
junction. The structural change could, thus, easily occur for the Au/H$_{2}$/Au junction due to smaller 
$E_{AC}$. The difference in the $dI/dV$ spectra between the Au/H$_{2}$/Au and Au/D$_{2}$/Au junction could be 
also explained by the number of the excitation process accompany the structural change. If $E_{AC}$ was 
larger than the energy of the Au-H$_{2}$ (D$_{2}$) vibration mode, the system should be excited into the states 
with higher vibrational quanta in a ladder climbing manner in order to overcome activation barrier 
for the structural change \cite{15}. In the multiple excitation process, the excitation of the system would 
be getting harder with the number of the excitation process. The number of the excitation process to 
overcome the activation barrier would be smaller for the Au/H$_{2}$/Au junction than that for the 
Au/D$_{2}$/Au junction due to the larger energy of the Au-H$_{2}$ vibration mode. Therefore, the structural 
change could easily occur for the Au/H$_{2}$/Au junction. The above discussion was supported by the 
previously reported result \cite{11}. Thijssen et al. reported that the "peak" spectra were observed for the 
Au contact in both H$_{2}$ and D$_{2}$ environment. They observed the "peak" spectra for the junction 
having a conductance value around 1 $G_{0}$. In the present study, the $dI/dV$ spectra were taken for the 
junction having a conductance value below 1 $G_{0}$ (around 0.1-0.4 $G_{0}$). 
It is because the clear difference in the conductance histogram between clean Au contact and Au contact in hydrogen environment 
was observed in the conductance regime around 0.1-0.4 $G_{0}$, and thus, the interaction between Au atomic contact and 
hydrogen was clear for the contact showing conductance around 0.1-0.4 $G_{0}$. 
The "peak" spectra could not 
be observed for the Au/D$_{2}$/Au junction. In the case of the vibrational heating mechanism, the 
reaction rate ($R$) per electron can be represented as $R \propto I^{n}$, where $I$ and $n$ are the current and the 
order of the reaction \cite{15}. The reaction rate shows nonlinear power-law dependence on current. The 
structural change would occur for the Au/D$_{2}$/Au junction having a high conductance value, while it 
would not occur for those having a low conductance value. These experimental results suggested 
that the structural change in the Au/D$_{2}$/Au junction would be the multiple excitation process.

\section{CONCLUSIONS}
\label{sec4}

The formation of the single Au/H$_{2}$/Au and Au/D$_{2}$/Au junction was revealed by the conductance 
measurement and action spectroscopy. In the $dI/dV$ spectra, symmetric peaks were observed for the 
Au/H$_{2}$/Au junction, while symmetric peaks were not observed for the Au/D$_{2}$/Au junction. The peaks 
in the $dI/dV$ spectra originated from the structural change induced by the phonon excitation. In the 
case of the Au/H$_{2}$/Au junction, the structural change induced by the phonon excitation could easily 
occur because the activation energy for the structural change and the number of excitation process 
would be small due to the large zero point energy and large energy of the Au-H$_{2}$ vibration mode.

\section{ACKNOWLEDGMENTS}
We would like to express our sincere gratitude to Dr. H. Nakamura and Dr. T. Tada at Tokyo University for the many stimulating discussions 
This work was supported by a Grant-in-Aid for Scientific Research on Priority Areas "Electron 
transport through a linked molecule in nano-scale" from MEXT.

\newpage

\end{document}